\documentclass[aps,prd,letterpaper,eqsecnum,nofootinbib]{revtex4}
\usepackage{verbatim}
\usepackage{amssymb}
\usepackage{amsmath}
\usepackage{amsfonts}
\usepackage{bm}

\begin{document}

\newcommand{\nin}{\noindent}
\newcommand{\nnm}{\nonumber}
\newcommand{\doe}{\partial}
\newcommand{\be}{\begin{equation}}
\newcommand{\ee}{\end{equation}}
\newcommand{\bea}{\begin{eqnarray}}
\newcommand{\eea}{\end{eqnarray}}
\newcommand{\bdm}{\begin{displaymath}}
\newcommand{\edm}{\end{displaymath}}
\newcommand{\bse}{\begin{subequations}}
\newcommand{\ese}{\end{subequations}}
\newcommand{\tb}{\textbf}
\newcommand{\n}{\,^{\scriptscriptstyle \text{n}}\!\;\!}
\newcommand{\pn}{\,^{\scriptscriptstyle \text{pn}}\!\;\!}
\newcommand{\cm}{\,^{\scriptscriptstyle \text{cm}}\!\;\!}
\newcommand{\bu}{\nin$\bullet\:\:$}
\newcommand{\mc}{\mathcal}
\newcommand{\gA}{{{\rm{g}},A}}
\newcommand{\lt}{\hat{\lambda}}
\newcommand{\ct}{x}
\newcommand{\oO}{+O(c^{-4})+O(\lambda^2)}

\title{Post-1-Newtonian tidal effects in the gravitational waveform from binary inspirals}

\author{Justin Vines}
\affiliation{Center for Radiophysics and Space Research, Cornell University, Ithaca, NY 14853, USA}

\author{\'{E}anna \'{E}. Flanagan}
\affiliation{Center for Radiophysics and Space Research, Cornell University, Ithaca, NY 14853, USA}

\author{Tanja Hinderer}
\affiliation{Theoretical Astrophysics, California Institute of Technology, Pasadena, CA 91125, USA}

\begin{abstract}

The gravitational wave signal from an inspiralling binary neutron star
system will contain detailed information about tidal coupling in the
system, and thus, about the internal physics of the neutron stars.  To
extract this information will require highly accurate models for the
gravitational waveform.  We present here a calculation of the
gravitational wave signal from a binary with quadrupolar tidal
interactions which includes all post-1-Newtonian-order effects in both
the conservative dynamics and wave generation.  We consider stars with
adiabatically induced quadrupoles moving in circular orbits, and work
to linear in the stars' quadrupole moments.  We find that
post-1-Newtonian corrections increase the tidal signal by
approximately 20\% at gravitational wave frequencies of 400 Hz.

\end{abstract}

\maketitle

\section{Introduction}

\subsection{Background and motivation}

Inspiralling and coalescing binary neutron stars are key sources for
ground-based gravitational wave (GW) detectors \cite{CutlerThorne}.
An important science goal in the detection of such sources is to
obtain robust information on the highly uncertain equation of state
(EoS) of neutron star matter \cite{EOS}.  The effects of the EoS on
the GW signal are largest during the late inspiral and merger stages
of binary evolution, at GW frequencies $\gtrsim 500$ Hz, and the
strong gravity and complex hydrodynamics involved in these regimes
require the use of fully relativistic numerical simulations for their
study (see e.g.~Ref.~\cite{Duez} and references therein).  A small but
clean EoS signature will also be present in the early inspiral
waveform, at frequencies $\lesssim500$ Hz within LIGO's most sensitive
band, arising from the effects of tidal coupling \cite{FH}.  The
relative weakness of orbital gravity in this regime makes it possible
to construct good approximate waveforms using post-Newtonian-based
analytic models \cite{BlanchetLRR}.

For point-particle models of binary inspiral, analytic gravitational
waveforms have been computed to 3PN accuracy \cite{3p5}, and spin
effects have been computed to 2PN accuracy \cite{2PNspin}.\footnote{
The shorthand $n$PN, for post-$n$-Newtonian, is used to describe corrections of order $c^{-2n}$ relative to Newtonian gravity, where $c$ is the speed of light.}
More recent efforts to improve the analytic description of neutron
star binary GW signals by including tidal effects began with
Refs.~\cite{FH,H}, which used a leading-order model of the tidal
coupling and GW emission to demonstrate the potential feasibility of
measuring EoS effects in inspiralling neutron stars in the low
frequency ($\lesssim 400$ Hz) regime with Advanced LIGO.  The tidal
contribution to the GW signal computed in Ref.~\cite{FH} depends on a
single tidal deformability parameter $\lambda$, which characterizes
the star's deformation response to a static (or adiabatically
changing) tidal field and which is sensitive to the star's EoS.

The quadrupolar tidal deformability $\lambda$ was defined in a fully
relativistic context and calculated for a variety of EoS models in
Refs.~\cite{FH,H,Hetal,DN1,BP}, and Refs.~\cite{DN1,BP} extended the
analysis to include higher-multipolar tidal responses of both
electric- and magnetic-type.  It was found in Ref.~\cite{Hetal} that
Advanced LIGO should be able to constrain the neutron stars' tidal
deformability to $\lambda\lesssim (1.2\times
10^{37}\,\rm{g}\,\rm{cm}^2\,\rm{s}^2)(D/100\,\rm{Mpc})$ with 95\%
confidence, for a binary of two 1.4 $M_\odot$ neutron stars at a
distance $D$ from the detector, using only the portion of the signal
with GW frequencies less than 400 Hz. The calculations of $\lambda$
for a 1.4 $M_\odot$ neutron star in Refs.~\cite{H,Hetal,DN1,BP}, using
several different equations of state, give values in the range
0.03--1.0$\times 10^{37}\,\rm{g}\,\rm{cm}^2\,\rm{s}^2$, so nearby
events may allow Advanced LIGO to place useful constraints on
candidate equations of state.

To detect or constrain the tidal deformability $\lambda$ will require
models for the tidal contribution to the GW signal that are accurate
to $\lesssim$10\%, much less than the current uncertainty in
$\lambda$.  References \cite{FH,Hetal} estimate the fractional
corrections to the tidal signal at GW frequencies below 400 Hz due to
several effects neglected by the model of the GW phasing used in
Ref.~\cite{FH}, namely, non-adiabaticity ($\lesssim$1\%),
higher-multipolar tidal coupling ($\lesssim$0.7\%), nonlinear
hydrodynamic effects ($\lesssim$0.1\%), spin effects
($\lesssim$0.3\%), nonlinear response to the tidal field
($\lesssim$3\%), viscous dissipation (negligible), and post-Newtonian
effects ($\lesssim$10\%).  The largest expected corrections, from
post-Newtonian effects in the orbital dynamics and GW emission, are
thus essential for an accurate analysis of the tidal signal.  These
corrections will depend on the neutron star physics only through the
same tidal deformability parameter $\lambda$ used in the Newtonian
treatment and thus can be easily incorporated into the same data
anaysis methods used in the Newtonian (tidal) case.

The extension of the tidal signal calculation to 1PN order was
recently discussed in Ref.~\cite{DN2} by Damour and Nagar (DN).
Working within the framework of the effective-one-body (EOB)
formalism, DN gave a complete description of the 1PN conservative
dynamics of tidally interacting binaries in circular orbits,
parametrized the forms of further 1PN corrections to the GW emission,
and made comparisons with numerical simulations (see also
Ref.~\cite{Baiotti}).  The 1PN conservative dynamics has also been
recently studied in Ref.~\cite{VF} by Vines and Flanagan (VF).
Working from the formalism for 1PN celestial mechanics developed in
Refs.~\cite{DSX1,DSX2} and extended by Ref.~\cite{RF}, VF found the
explicit equations of motion and action principle for generic orbits
and generic evolution of the bodies' quadrupoles.  Specializing to
adiabatically induced quadrupoles and circular orbits, the results of
VF agree with those of DN for the 1PN conservative dynamics.  The
construction of the 1PN metric given by VF also allows for explicit
computation of the binary system's 1PN-accurate mass multipole
moments.

In the present paper, we use the results of VF \cite{VF} to derive the
1PN-accurate GW signal from an inspiralling binary with quadrupolar
tidal interactions.  Working to linear order in the stars' quadrupole
moments, and using adiabatically induced quadrupoles and circular
orbits, we compute the binary's binding energy and GW energy flux and
use them to determine the phase evolution of the emitted GW signal in
the stationary phase approximation.  The results presented here can be
used to extend the validity of analytic GW signals to higher
frequencies, and to provide useful information for hybrid schemes that
attempt to bridge the gap in frequencies between analytic inspiral
models and the start of numerical simulations, such as the EOB
formalism of Ref.~\cite{DN2}.  Our expressions for the orbital
equations of motion and binding energy may also be useful for the
construction of quasi-equilibrium initial data for numerical
simulations \cite{initialdata}.  We note that the 1PN corrections
calculated here slightly improve the prospects for detection of tidal
effects in binary GW signals, as they increase the tidal signal by
$\sim20\%$ at GW frequencies of 400 Hz.

\subsection{Organization}

The organization of this paper is as follows.  In Sec.~\ref{sec:cons},
we briefly state the key results of Ref.~\cite{VF} for the 1PN
conservative dynamics of a binary in which one member has a mass
quadrupole moment.  We specialize to the adiabatic limit and circular
orbits and compute the gauge-invariant binding energy as a function of
orbital frequency. In Sec.~\ref{sec:fluxes}, we consider the
gravitational radiation and obtain the 1PN tidal corrections to the
radiated energy flux.  We then compute the resulting 1PN tidal
corrections to the phase of the Fourier transform of the waveform in
the stationary phase approximation and conclude in Sec.~\ref{disc}
with a short discussion of the results.

\subsection{Notation and conventions}

We use units where Newton's constant is $G=1$, but retain factors of
the speed of light $c$, with $1/c^2$ serving as the formal expansion
parameter for the post-Newtonian expansion.  We use lowercase latin
letters $a,b,i,j,\ldots$ for indices of spatial tensors.  Spatial
indices are contracted with the Euclidean metric,
$v^iw^i=\delta_{ij}v^iw^j$, with up or down placement of the indices
having no meaning.
We use angular brackets to denote the symmetric, trace-free projection
of tensors, for example
$T^{<ab>}=T^{(ab)}-\frac{1}{3}\delta^{ab}T^{cc}$.

\section{Conservative dynamics in the adiabatic limit}\label{sec:cons}

In this section we briefly review the key results of VF \cite{VF}
concerning the 1PN conservative dynamics of a binary system with
quadrupolar tidal coupling.  For simplicity, we consider a binary
composed of one point-mass (body 1) and one deformable star (body 2).
Since we consistently work to linear order in the quadrupole, our
results can be easily generalized to the case of two deformable bodies
by interchanging body labels.  The binary's orbital dynamics can be
formulated in terms of the separation (three-)vector $z^i=z^i_2-z^i_1$
between the bodies, the bodies' masses $M_1$ and $M_2$, and the
quadrupole moment $Q_2^{ij}$ of body 2.

The 1PN-accurate worldlines $x^i=z_1^i(t)$ and $x^i=z_2^i(t)$ of the
bodies' centers of mass-energy and their separation
$z^i(t)=z_2^i(t)-z_1^i(t)$ are defined in a `global' 1PN coordinate
system $(t,x^i)$.  The global coordinates are conformally Cartesian
and harmonic, and they tend to inertial coordinates in Minkowski
spacetime as $|\bm x|\to\infty$.  Also, the binary system's center of
mass-energy is taken to be at rest at the origin $x^i=0$ (the system's
1PN-accurate mass dipole moment is set to zero), so that the $(t,x^i)$
coordinates correspond to the center-of-mass-energy frame of the
system.   We use the following notation for the relative position,
velocity, and acceleration:
\bea
&z^i=z_2^i-z_1^i,\quad r=|\bm{z}|=\sqrt{\delta_{ij}z^iz^j},\quad n^i=z^i/r,&
\nnm\\\nnm
&v^i=\dot z^i,\quad \dot r=v^in^i,\quad a^i=\ddot z^i,&
\eea
with dots denoting derivatives with respect $t$.

We take $M_1$ and $M_2$ to be the bodies' conserved rest
masses,\footnote{Note that the mass $M_2$ used here is not the
  1PN-accurate Blanchet-Damour \cite{BD} mass monopole moment (which
  was called $M_2$ in VF \cite{VF}); rather, the $M_2$ used here is
  the conserved part of the BD mass monopole (called $\n M_2$ in VF
  \cite{VF}).  The full 1PN-accurate monopole also receives
  contributions from the body's internal elastic energy (and from the
  tidal gravitational potential energy), which for a deformable body,
  will vary as tidal forces do work on the body.  The effects of these
  time-dependent contributions to the monopole have been separately
  accounted for in the Lagrangian (\ref{Lad}), and the mass $M_2$
  appearing there is constant.}
and we define the total mass $M$, mass fractions $\chi_1,\chi_2$,
reduced mass $\mu$, and symmetric mass ratio $\eta$ by
\be
M=M_1+M_2,\quad \chi_1=M_1/M,\quad \chi_2=M_2/M,\quad\mu=\eta M=\chi_1\chi_2 M.
\ee
Note that there are only two independent parameters among these
quantities; we will tend to express our results in terms of the total
mass $M$ and the mass fraction $\chi_2$ of the deformable body, unless
factorizations make it more convenient to use $\chi_1=1-\chi_2$ or
$\eta=\chi_1\chi_2$.

The tidal deformation of body 2 is described by its 1PN-accurate
Blanchet-Damour \cite{BD} mass quadrupole moment $Q_2^{ij}(t)$.  We
will work in the limit where the quadrupole is adiabatically induced
by the tidal field; i.e.~we assume that the quadrupole responds to the
instantaneous tidal field according to
\bse\label{pnQG}
\be\label{pnQ}
Q_2^{ij}(t)=\lambda G_2^{ij}(t).
\ee
Here, the constant $\lambda$ is the tidal deformability,\footnote{The
  tidal deformability is related to the Love number $k_2$ \cite{Love}
  and the star's areal radius $R$ by $\lambda=2k_2R^5/3$.}
and $G_2^{ij}(t)$ is the quadrupolar gravito-electric DSX \cite{DSX2}
tidal moment of body 2 which encodes the leading order ($l=2$) tidal
field felt by body 2.  For the binary system under consideration, the
tidal moment is given by
\bea\label{pnG}
G_2^{ij}&=&\frac{3\chi_1 M}{r^3}n^{<ij>}
+\frac{1}{c^2}\frac{3\chi_1 M}{r^3}
\bigg[\left(2v^2-\frac{5\chi_2^2}{2}\dot{r}^2-\frac{6-\chi_2}{2}\frac{M}{r}\right)n^{<ij>}
+v^{<ij>}-(3-\chi_2^2)\dot{r}n^{<i}v^{j>}\bigg]
\nnm\\
&&+O(c^{-4})+O(\lambda).
\eea
\ese

With the quadrupole given by Eqs.~(\ref{pnQG}) in the adiabatic limit,
the only independent degree of freedom is the binary's relative
position $z^i(t)$.  It was shown by VF \cite{VF} that the evolution of
$z^i(t)$ is governed by the Lagrangian
\bea
\mc L[z^i]&=&\frac{\mu v^2}{2}+\frac{\mu M}{r}\left(1+\frac{\Lambda}{r^5}\right)
+\frac{\mu}{c^2}\left\{\theta_0v^4
+\frac{M}{r}\left[v^2\left(\theta_1+\xi_1\frac{\Lambda}{r^5}\right)
+\dot{r}^2\left(\theta_2+\xi_2\frac{\Lambda}{r^5}\right)
+\frac{M}{r}\left(\theta_3+\xi_3\frac{\Lambda}{r^5}\right)\right]\right\}
\nnm\\
&&\oO,\label{Lad}
\eea
with $\Lambda=(3\chi_1/2\chi_2)\lambda$, and with the dimensionless
coefficients
\bea\label{Ladcs}
&\theta_0=(1-3\eta)/8,
\quad
\theta_1=(3+\eta)/2,
\quad
\theta_2=\eta/2,
\quad
\theta_3=-1/2,&
\nnm\\
&\xi_1=(\chi_1/2)(5+\chi_2),
\quad
\xi_2=-3(1-6\chi_2+\chi_2^2),
\quad
\xi_3=-7+5\chi_2.&
\eea
The orbital equation of motion resulting from this Lagrangian, via
$(d/dt)(\doe\mc L/\doe v^i)=\doe\mc L/\doe z^i$, is given by
\bea\label{ada}
a^i&=&-\frac{Mn^i}{r}\left(1+\frac{6\Lambda}{r^5}\right)
+\frac{M}{c^2r^2}\left[
v^2n^i\left(\phi_1+\zeta_1\frac{\Lambda}{r^5}\right)
+\dot{r}^2n^i\left(\phi_2+\zeta_2\frac{\Lambda}{r^5}\right)
+\frac{M}{r}n^i\left(\phi_3+\zeta_3\frac{\Lambda}{r^5}\right)
+\dot{r}v^i\left(\phi_4+\zeta_4\frac{\Lambda}{r^5}\right)\right]
\nnm\\
&&\oO,
\eea
with coefficients
\bea
&\phi_1=-1-3\eta,
\quad
\phi_2=3\eta/2,
\quad
\phi_3=2(2+\eta),
\quad
\phi_4=2(2-\eta),&
\nnm\\
&\zeta_1=-3(2-\chi_2)(1+6\chi_2),
\quad
\zeta_2=24(1-6\chi_2+\chi_2^2),
\quad
\zeta_3=66+9\chi_2-19\chi_2^2,
\quad
\zeta_4=6(2-\chi_2)(3-2\chi_2).&
\eea
The conserved energy constructed from the Lagrangian (\ref{Lad}) is
\bea\label{Ead}
E&=&v^i\doe\mc L/\doe v^i-\mc L
\nnm\\
&=&\frac{\mu v^2}{2}-\frac{\mu M}{r}\left(1+\frac{\Lambda}{r^5}\right)
+\frac{\mu}{c^2}\left\{3\theta_0v^4
+\frac{M}{r}\left[v^2\left(\theta_1+\xi_1\frac{\Lambda}{r^5}\right)
+\dot{r}^2\left(\theta_2+\xi_2\frac{\Lambda}{r^5}\right)
-\frac{M}{r}\left(\theta_3+\xi_3\frac{\Lambda}{r^5}\right)\right]\right\}
\nnm\\
&& \oO,
\eea
which is a constant of motion of the equation of motion (\ref{ada}).

The orbital equation of motion (\ref{ada}) admits solutions of the form
\bse\label{circ}
\be
z^i(t)=rn^i(t)=r(\cos(\omega t),\sin(\omega t),0),
\ee
with $ \dot r=0$, $v^2=r^2\omega^2$ and $a^i=-r\omega^2 n^i$,
corresponding to circular orbits in the $x$-$y$ plane with frequency
$\omega$.  For later convenience, we introduce the unit vector
$\phi^i$ in the direction of the velocity $v^i$, which satisfies
\be
\dot z^i=v^i=r\omega\phi^i,\qquad \dot n^i=\omega \phi^i,\qquad \dot\phi^i=-\omega n^i,\qquad n^i\phi^i=0,
\ee
\ese
for circular orbits.  Working to linear order both in the
post-Newtonian parameter $c^{-2}$ and in the tidal deformability
parameter $\lambda$, Eqs.~(\ref{ada}) and (\ref{circ}) yield the
radius-frequency relationship
\be\label{rofomega}
r(\omega)=\frac{M^{1/3}}{\omega^{2/3}}\left[1+\frac{3\chi_1}{\chi_2}\lt
+\frac{\eta-3}{3}\ct
+\frac{\chi_1}{2\chi_2}\left(-6+26\chi_2-\chi_2^2\right)\ct\lt \right]\oO.
\ee
Here, we have introduced the $\omega$-dependent dimensionless quantities
\be\label{lhx}
\lt\equiv\frac{\lambda\omega^{10/3}}{M^{5/3}},\qquad\qquad\ct\equiv\frac{(M\omega)^{2/3}}{c^2},
\ee
which characterize the fractional corrections due to tidal effects and
to post-Newtonian effects.  Using Eqs.~(\ref{Ead}), (\ref{circ}) and
(\ref{rofomega}), we can also find the gauge-invariant
energy-frequency relationship for circular orbits:
\bea\label{eofomega}
E(\omega)&=&\mu(M\omega)^{2/3}\bigg[-\frac{1}{2}
+\frac{9\chi_1}{2\chi_2}\lt
+\frac{9+\eta}{24}\ct
+\frac{11\chi_1}{4\chi_2}(3+2\chi_2+3\chi_2^2)\ct\lt\bigg]\oO.\phantom{yoy}
\eea
This expression for the binding energy can be directly compared with
Eqs.~(37,38,50-57) of DN \cite{DN2}, and indicates that their
parameter $\bar\alpha_1'$ giving the 1PN tidal contribution to the
binding energy should have the value
$\bar\alpha_1'=(11/18)(3+2\chi_2+3\chi_2^2)$ instead of $55\chi_2/18$
(note the the quantity denoted here by $\chi_2$ is denoted by $X_A$ in
DN \cite{DN2}).  For the case of equal masses ($\chi_1=\chi_2=1/2$,
$\eta=1/4$), the binding energy (\ref{eofomega}) simplifies to
\be
E_{M_1=M_2}(\omega)=-\frac{M^{5/3}\omega^{2/3}}{8}\left[1-\frac{37}{48}\ct
-18\lt\left(1+\frac{209}{72}\ct\right)\right]\oO.\label{equalmeb}
\ee
For orbital frequencies of 200 Hz (GW frequencies of 400 Hz) and total
mass $M=2.8 M_\odot$, the 1PN fractional correction to the Newtonian
tidal term in the binding energy is $(209/72)\ct \approx 19\%$.

\section{Gravitational radiation}\label{sec:fluxes}

The energy flux from the binary due to gravitational radiation is
determined by the time variation of the binary system's multipole
moments \cite{BlanchetLRR}.  The flux $\dot E$ to 3.5PN-order (or to
1PN-order relative to the leading 2.5PN flux) is given in terms of the
total system's mass quadrupole moment $Q_{\rm{sys}}^{ij}(t)$, current
quadrupole moment $S_{\rm{sys}}^{ij}(t)$, and mass octupole moment
$Q_{\rm{sys}}^{ijk}(t)$ by
\be
\dot E=-\frac{1}{5c^5} ( \doe_t^3 Q_{\rm{sys}}^{ij} )^2
-\frac{1}{c^7}\left[\frac{1}{189}(\doe_t^4 Q_{\rm{sys}}^{ijk})^2
+\frac{16}{45} (\doe_t^3 S_{\rm{sys}}^{ij})^2 \right] + O(c^{-8}),
\label{eq:fluxformula}
\ee
c.f.~Eq.~(223) of Ref.~\cite{BlanchetLRR}.

The binary system's multipole moments can be computed from the
asymptotic form of the global metric, as in Sec.~IV of Ref.~\cite{VF}.
The mass quadrupole $Q_{\rm{sys}}^{ij}$, which is needed to 1PN
accuracy in the flux formula (\ref{eq:fluxformula}), can be found from
Eqs.~(4.6,4.5,B4,B5,6.1) of VF \cite{VF}; the result is
\bea
Q^{ij}_{\rm{sys}}&=&Q^{ij}_2+\mu r^2 n^{<ij>} +\frac{\mu r^2}{c^2}\bigg\{
n^{<ij>}\left[ v^2 \left(\tau_1+\sigma_1\frac{\lambda}{r^5}\right) + \dot{r}^2 \left(\tau_2+\sigma_2\frac{\lambda}{r^5}\right)
+ \frac{M}{r} \left(\tau_3+\sigma_3\frac{\lambda}{r^5}\right) \right]
\nnm\\
&&\phantom{Q^{ij}+\mu r^2 n^{<ij>} +\frac{\mu r^2}{c^2}\bigg\{}
+ v^{<ij>} \left(\tau_4+\sigma_4\frac{\lambda}{r^5}\right) + \dot{r} n^{<i}v^{j>} \left(\tau_5+\sigma_5\frac{\lambda}{r^5}\right) \bigg\} \oO, \label{eq:qijt}
\eea
where the 1PN-accurate body quadrupole $Q_2^{ij}$ is given by
Eqs.~(\ref{pnQG}) above and the dimensionless coefficients  $\tau$ and
$\sigma$ are given by
\bea\label{tau4}
\tau_1&=&\frac{29}{42}(1-3\eta),\quad \tau_2=0,\quad \tau_3=\frac{1}{7}(8\eta-5),\quad \tau_4=\frac{11}{21}(1-3\eta),\quad \tau_5=\frac{4}{7}(3\eta-1),
\\ \nnm
\sigma_1&=&\frac{13\chi_1^2}{7\chi_2},\quad \sigma_2=\frac{185\chi_1^2}{14\chi_2},\quad \sigma_3=-\frac{3\chi_1}{14\chi_2}(8+23\chi_2+13\chi_2^2),\quad
\sigma_4=\frac{38\chi_1^2}{7\chi_2},\quad \sigma_5=-\frac{151\chi_1^2}{7\chi_2}.
\eea
This result holds for generic orbits (in a binary where body 2 has an
adiabatically induced quadrupole).
Using Eqs.~(\ref{pnQG}) for the body quadrupole, Eqs.~(\ref{circ}) to
specialize to circular orbits, and the radius-frequency relationship
(\ref{rofomega}), the system quadrupole simplifies to
\be
Q^{ij}_{\rm{sys}}=\frac{\eta M^{5/3}}{\omega^{4/3}}\left[n^{<ij>}(1+\sigma_0\lt)
+\ct\left(\tau_6 n^{<ij>} + \tau_4 \phi^{<ij>} \right)+\ct\lt\left(\sigma_6 n^{<ij>} + \sigma_7 \phi^{<ij>} \right)\right]\oO,\label{Qsysf}
\ee
with $\tau_4$ as in Eq.~(\ref{tau4}), and with
\bdm
\sigma_0=\frac{3(3-2\chi_2)}{\chi_2},\:\tau_6=-\frac{85+11\eta}{42},\:\sigma_6=\frac{1}{14\chi_2}(4+56\chi_2+264\chi_2^2-219\chi_2^3),\:
\sigma_7=\frac{1}{7\chi_2}(103-252\chi_2+302\chi_2^2-132\chi_2^3).
\edm
The expression (\ref{Qsysf}) for the total quadrupole determines the
unknown 1PN correction coefficient introduced in Eq.~(71) of DN
\cite{DN2}.\footnote{The parametrization of the tidal contribution to
  the system quadrupole given in Eqs.~(68-71) of DN \cite{DN2} does
  not quite match the form given in Eq.~(\ref{Qsysf}) here, as no
  $\phi^{<ij>}$ term is included.  Also, their parametrization leaves
  some dependence on the radius $r$, while ours eliminates $r$ in
  favor of the gauge invariant quantity $\omega$.  Still, as the
  coefficients of $\ct\lt n^{<ij>}$ and $\ct\lt \phi^{<ij>}$ in our
  Eq.~(\ref{Qsysf}) end up additively combined in the final
  contribution to the energy flux, one could in principle determine an
  effective value for the coefficient $\beta_1$ in Eq.~(71) of DN
  \cite{DN2} that would lead to the correct flux $\dot E$.}

Similarly, the system's mass octupole and current quadrupole, which
are needed only to Newtonian order, are given by
\bea
Q^{ijk}_{\rm{sys}}&=&\mu r^3 n^{<ijk>}\left[(\chi_1-\chi_2)+\frac{9\chi_1}{\chi_2}\frac{\lambda}{r^5}\right]+O(c^{-2})+O(\lambda^2)
\nnm\\
&=&\frac{\eta M^2}{\omega^2}n^{<ijk>}\left[(\chi_1-\chi_2)+\frac{18\chi_1^2}{\chi_2}\lt\right]+O(c^{-2})+O(\lambda^2),
\label{eq:qijkt}
\eea
and
\bea
S^{ij}_{\rm{sys}}&=&\mu r^2 \epsilon^{kl<i}n^{j>k}v^l\left[(\chi_1-\chi_2)+\frac{9\chi_1}{2\chi_2}\frac{\lambda}{r^5}\right]+O(c^{-2})+O(\lambda^2)
\nnm\\
&=&\frac{\eta M^2}{\omega}\epsilon^{kl<i}n^{j>k}\phi^l\left[(\chi_1-\chi_2)+\frac{9\chi_1(3-4\chi_2)}{2\chi_2}\lt\right]+O(c^{-2})+O(\lambda^2),
\label{eq:sijt}
\eea
where the first equalities hold for generic orbits, and the second
equalities hold for circular orbits.

Having gathered the expressions (\ref{Qsysf}), (\ref{eq:qijkt}) and
(\ref{eq:sijt}) for the system multipole moments, we can insert them
into the flux formula (\ref{eq:fluxformula}).  Using also
Eqs.~(\ref{circ}) to for the time derivatives of $n^i$ and $\phi^i$
(which are the only time-dependent quantities in the final expressions
for the multipoles), and working out the STF projections and
contractions (e.g.~$n^{<ijk>}n^{<ijk>}=2/5$) using the STF identities
from (e.g.) Ref.~\cite{RF}, we find the GW energy flux from the binary
to be given by
\bea
\dot E(\omega)&=&-\frac{32}{5}\eta^2\ct^{\,5/2}
\bigg[1-\left(\frac{1247}{336}+\frac{35\eta}{12}\right)\ct+\frac{6(3-2\chi_2)}{\chi_2}\lt+\frac{1}{28\chi_2}\left(-704-1803\chi_2+4501 \chi_2^2 -2170\chi_2^3\right)\ct\lt
\nnm\\
&&+O(c^{-3})+O(\lambda^2)\bigg].\label{flux}
\eea
The coefficients for the 1PN point-mass (second) and Newtonian tidal
(third) terms match those given in Refs.~\cite{BlanchetLRR,FH}.

Using energy balance and the stationary phase approximation
\cite{TichyPhase}, the Fourier transform of the gravitational waveform
can be written as
$h={\cal A}e^{i\psi}$, with the phase $\psi(\omega)$ determined from
the binding energy $E(\omega)$ and flux $\dot E(\omega)$ as functions
of the orbital frequency $\omega$ by the relation
\be
\frac{d^2\psi}{d\omega^2}=\frac{2}{\dot E}\frac{dE}{d\omega}.
\ee
Taking $\dot E$ from Eq.~(\ref{flux}), finding $dE/d\omega$ from a derivative of Eq.~(\ref{eofomega}), and integrating twice (dropping unimportant integration constants) yields the phase:
\bea
\psi(\omega)&=&\frac{3}{128\eta \ct^{\,5/2}}\left[1+\psi_{0,1}\lt+\psi_{1,0}\ct+\psi_{1,1}\ct\lt+O(c^{-3})+O(\lambda^2)\right]
\\ \nnm
&=&\frac{3c^5}{128\eta(M\omega)^{5/3}}\left[1+\psi_{0,1}\frac{\lambda\omega^{10/3}}{M^{5/3}}+\psi_{1,0}\frac{(M\omega)^{2/3}}{c^2}
+\psi_{1,1}\frac{\lambda\omega^4}{Mc^2}+O(c^{-3})+O(\lambda^2)\right],
\eea
with coefficients
\be
\psi_{0,1}=-\frac{24}{\chi_2}(1+11\chi_1),\quad
\psi_{1,0}=\frac{20}{9}\left(\frac{743}{336}+\frac{11\eta}{4}\right),\quad
\psi_{1,1}=-\frac{5}{28\chi_2}\left(3179-919\chi_2-2286\chi_2^2+260\chi_2^3\right).
\ee

The above results concern a binary where only one body (body 2)
develops a tidally induced quadrupole, with quadrupolar tidal
deformability $\lambda_2=\lambda$.  For the case of two deformable
bodies, the contribution to the tidal signal from the other body (body
1) can simply be added to the phase by interchanging body labels
$(1\leftrightarrow2)$ in the tidal terms.  For the case of equal
masses and identical equations of state,
$M_1=M_1=M/2$ and $\lambda_1=\lambda_2=\lambda$, the phase correction is
\be
\psi_{M_1=M_2}(\omega)=\frac{3}{32\ct^{\,5/2}}\left[1-624\lt+\frac{2435}{378}\ct-\frac{3115}{2}\ct\lt \right].
\ee
The 1PN correction increases the tidal signal by $\approx 17\%$ at
gravitational wave frequencies of $400$Hz for $M=2.8 M_\odot$.

\bigskip

From the expressions (\ref{flux}) and (\ref{eofomega}) for the
gravitational wave luminosity $\dot E(\omega)$ and the binding energy
$E(\omega)$, it is straightforward to construct the phase $\varphi(t)$
of the time-domain gravitational waveform based on the various PN
Taylor approximants used in several approaches to interfacing
analytical and numerical relativity \cite{templatecompare}.  We
provide here the explicit expressions for the Taylor T4 approximant,
in which the function ${\cal F}\equiv \dot E/(dE/dx)$ is expanded in a
Taylor series and the differential equations
\be
\frac{dx}{dt}={\cal F}, \ \ \ \ \ \frac{d\varphi}{dt}=2 x^{3/2}/M,
\ee
are integrated numerically [with $x$ as in (\ref{lhx})]. The tidal
contribution to the function ${\cal F}^{\rm T4}$ adds linearly to the
$3.5$PN point mass terms and is given to 1PN order by
\be
{\cal F}^{\rm{T4}}_{\rm{tidal}}=
\frac{32\chi_1\lambda_2}{5M^6}\left[12(1+11\chi_1)x^{10}
+\left(\frac{4421}{28}-\frac{12263}{28}\chi_2+\frac{1893}{2}\chi_2^2-661\chi_2^3\right)x^{11}\right]+(1\leftrightarrow 2).
\ee

\section{Discussion and Conclusions}\label{disc}
We have provided the 1PN accurate description of quasi-circular binary
inspiral with quadrupolar tidal coupling and obtained the 1PN tidal
contributions to the phasing of the emitted gravitational radiation in
the low-frequency, adiabatic limit.  Our results show that 1PN effects
increase the tidal corrections by approximately $20\%$ at
gravitational wave frequencies of $400$ Hz in the case of two
$1.4M_\odot$ stars.  These results should be of use in constructing GW
measurement templates and can be easily be incorporated into the EOB
formalism as discussed by DN \cite{DN2}; the unknown coefficients
introduced by DN pertaining to 1PN quadrupolar tidal effects have been
determined here.  Our results can also be of use in comparing
numerical and analytic waveforms and constructing initial data for
numerical simulations.  While we have restricted attention here to the
case of circular orbits, the results necessary to compute the GW
signal for generic orbits can all be found in this paper.  This work
could also be extended to consider 1PN tidal coupling at higher
multipolar orders; the necessary machinery (and the template of the
quadrupolar case) is fully contained in VF \cite{VF}.

\begin{acknowledgments}

This research was supported at Cornell by NSF Grant PHY-0757735, and at Caltech by
the Sherman Fairchild Foundation.

\end{acknowledgments}

\bibliography{bt}

\end{document}